\documentclass[12pt]{article}
\usepackage{amssymb,amsmath}
\newcommand{\be}{\begin{equation}}
\newcommand{\ee}{\end{equation}}
\newcommand{\beq}{\begin{eqnarray}}
\newcommand{\eeq}{\end{eqnarray}}
\newcommand{\ba}{\begin{array}}
\newcommand{\ea}{\end{array}}
\newcommand{\beqn}{\begin{eqnarray*}}
\newcommand{\eeqn}{\end{eqnarray*}}

\newcommand{\f}[2]{\frac{#1}{#2}}

\newcommand{\roughly}[1]{\raise.3ex\hbox{$#1$\kern-.75em
\lower1ex\hbox{$\sim$}}}

\begin{document}

\centerline{\Large\bf Supermassive black holes or boson stars?} 
\centerline{\Large\bf Hair counting with gravitational wave detectors}
\bigskip
\bigskip
\centerline{{\bf Emanuele Berti}
\footnote{email: berti@wugrav.wustl.edu}}
\smallskip
\centerline{McDonnell Center for the Space Sciences}
\centerline{Department of Physics, Washington University, St. Louis, Missouri 63130, USA}
\bigskip
\centerline{{\bf Vitor Cardoso}
\footnote{email: vcardoso@phy.olemiss.edu}}
\centerline{Dept. of Physics and Astronomy, The University of
Mississippi, University, MS 38677-1848, USA}
\bigskip
\bigskip
\bigskip

\begin{abstract}
The evidence for supermassive Kerr black holes in galactic centers is
strong and growing, but only the detection of gravitational waves will
convincingly rule out other possibilities to explain the
observations. The Kerr spacetime is completely specified by the first
two multipole moments: mass and angular momentum. This is usually
referred to as the ``no-hair theorem'', but it is really a
``two-hair'' theorem. If general relativity is the correct theory of
gravity, the most plausible alternative to a supermassive Kerr black
hole is a rotating boson star.  Numerical calculations indicate that
the spacetime of rotating boson stars is determined by the first {\it
three} multipole moments (``three-hair theorem''). The Laser
Interferometer Space Antenna ({\em LISA}) could accurately measure the
oscillation frequencies of these supermassive objects.  We propose to
use these measurements to ``count their hair'', unambiguously
determining their nature and properties.
\end{abstract}

\newpage

The present observational evidence for the existence of astrophysical
black holes (BHs) at the centers of galaxies is strong and growing
\cite{narayan}.  The most convincing case comes from observations of
stellar proper motion at the center of our own Galaxy, indicating the
presence of a ``dark object'' of mass $M\simeq (3.7\pm 0.2)\times
10^6~M_\odot$ \cite{SgrA} and size smaller than about one astronomical
unit \cite{shen}. A Schwarzschild BH of the given mass has radius
$R=2GM/c^2\simeq 0.073$ astronomical units, compatible with the
observations.

The massive BH picture has become a paradigm. All scientific paradigms
are dangerous, since it is easy to believe that we have proved what we
expect to find. To argue convincingly that galactic centers contain
BHs we must rule out all possible alternatives. Some of these
alternatives are already incompatible with observations. For example,
the central mass in our Galaxy (and several other galactic nuclei)
cannot be modelled by any distribution of individual objects: such a
distribution would be gravitationally unstable \cite{maoz}. A possible
exotic alternative to BHs (fermion balls) is also incompatible with
observations of the supermassive object at the Galactic center
\cite{schodel}.

Assessing the BH nature of these massive objects by electromagnetic
observations is a hard task.  Despite recent claims that (under
plausible assumptions) an event horizon is required to explain
observations of the massive object at the center of our own Galaxy
\cite{BN}, some hold the view that an observational proof of the
existence of event horizons based on electromagnetic observations is
fundamentally impossible \cite{AKL}.

A conclusive test of the nature of these supermassive objects will
come from observations of gravitational radiation with the Laser
Interferometer Space Antenna ({\em LISA}). Elaborating on ideas
proposed by Ryan \cite{ryanmap}, Kesden {\it et al.}  \cite{kesden}
showed that if the central object has no horizon (their particular
model assumes it is a soliton star) this will leave a characteristic
imprint on the gravitational radiation emitted by the inspiral of
solar-mass compact objects.  Ryan's proposal to map spacetimes using
inspiral waveforms is promising, but the data analysis task is plagued
by a ``confusion problem'': the possibility of misinterpreting truly
non-Kerr waveforms by Kerr waveforms with different orbital parameters
\cite{GB}.

Here we propose a different idea, based on measuring the oscillation
frequencies (``ringdown waves'') of isolated supermassive objects.  In
Ref.~\cite{BCW} we provided a general criterion (valid for any
interferometric gravitational wave detector) on the signal-to-noise
ratio required to test the BH nature of the source from a detection of
ringdown waves. In the next section we explain the basic idea behind
this test, summarizing our analysis in \cite{BCW}. 

Gravitational wave astronomy will look at the Universe from a new
perspective, and is likely to present us with surprises. For this
reason in the rest of the paper we speculate on possible alternatives
to the BH hypothesis, under the ``conservative'' assumption that
general relativity is the correct theory of gravity. We focus on the
most promising (and fascinating) family of particle-physics candidates
which have not been ruled out by observations: boson stars
\cite{torres,SM}.

\subsection*{Black hole ringdown and tests of the no-hair (two-hair) theorem}  

Newly formed BHs are expected to emit characteristic radiation
(ringdown waves) whose amplitude can be expressed as a linear
superposition of a discrete set of damped oscillations, known as
quasi-normal modes (QNMs):
\be h=h_++ih_\times=
\f{M}{r}\sum_{lmn} {\cal A}_{lmn} e^{i(\omega_{lmn}t+\phi_{lmn})}
S_{lmn}\,.
\ee

The functions $S_{lm}$ are spin-weighted spheroidal harmonics, used to
separate the angular dependence of the perturbations and characterized
by angular quantum numbers $(l,m)$.  The amplitudes ${\cal A}_{lmn}$
and phases $\phi_{lmn}$ of the various modes are determined by the
specific process that formed the final hole. 
The complex mode frequencies $\omega_{lmn}=2\pi f_{lmn}+i/\tau_{lmn}$
($f_{lmn}$ is the oscillation frequency and $\tau_{lmn}$ is the
damping time of the oscillation) are labelled by an integer $n$ that
sorts them by the magnitude of their imaginary part. Large imaginary
part means short damping time, so low-$n$ modes are (in principle)
easier to observe. According to the standard lore, $l=m=2$ modes
(having the longest damping time and corresponding to bar-shaped
deformations) should dominate the signal \cite{FH}.  
The QNM frequencies $\omega_{lmn}$ are universal, depending only on
the mass $M$ and specific angular momentum $a=J/M$ of the BH. This
universality is a consequence of the so-called ``no-hair'' theorem,
which is really a ``two-hair'' theorem: the spacetime of a Kerr BH (as
determined by the mass multipole moments $M_n$ and current multipole
moments $S_n$) is completely specified once we know the first two
moments $M_0=M$ and $S_1=J=Ma$, since $M_n+i S_n=M(ia)^n$.

{\em LISA} should be able to detect ringdown waves from oscillating
supermassive BHs throughout the observable universe in the frequency
range $(10^{-5}-10^{-1})$~Hz, with maximum sensitivity around
$10^{-2}$~Hz. The signal-to-noise ratio (SNR) depends on the BH's
distance, mass, angular momentum and (more importantly) on the
ringdown efficiency $\epsilon_{rd}$ (fraction of mass radiated in
ringdown waves). Usually the SNR is quite large, because ringdown
waves from supermassive BHs are emitted in the maximum sensitivity
range of {\em LISA}: the fundamental QNM of a Schwarzschild BH has
frequency and damping time
\be\label{BHs}
f_{200}= 1.2\cdot10^{-2} (10^6 M_\odot/M)~{\rm Hz}\,,
\qquad
\tau_{200}=55~(M/10^6 M_\odot)~{\rm s}\,.
\ee
For example, a BH with $a/M=0.8$ and mass
$\sim 4\times 10^6~M_\odot$ that radiates 1~\% of its mass at
luminosity distance $D_L=3$~Gpc will be detected by {\em LISA} with
SNR $\sim 10^4$ (see Fig.~8 of \cite{BCW}). Since measurement errors
are inversely proportional to the SNR, typical measurements of QNM
frequencies should be exquisitely precise.

Detection of a single QNM provides us with two observables: $f_{lmn}$
and $\tau_{lmn}$, which are functions of $M$ and $a$ only.  Inverting
these (known) functional relations we can determine both the mass and
the angular momentum of the BH, {\it if} we know which mode we are
detecting. In order to test the no-hair (``two-hair'') theorem,
however, it is necessary (though not sufficient) to resolve two
QNMs. Roughly speaking, one mode is used to measure $M$ and $a$, the
other to test consistency with the GR prediction.

In \cite{BCW}, using a simple extension of the Rayleigh criterion for
resolving sinusoids, we estimated the SNR required to resolve the
frequencies and/or damping times of various pairs of modes, as a
function of $a/M$. Under the plausible assumption that the amplitude
of the first overtone ($n=1$) is $1/10$ that of the fundamental mode
($n=0$) we found that tests of the no-hair theorem should be feasible,
even under rather pessimistic assumptions on $\epsilon_{rd}$, as long
as the first overtone radiates a fraction $\sim 10^{-2}$ of the total
ringdown energy.  However, resolving {\em both} frequencies and
damping times typically requires a SNR greater than about $10^3$.
This is only possible under rather optimistic assumptions about the
radiative efficiency, and it can be impossible if the dominant mode
has $l=m=2$ and the BH is rapidly spinning.  Requiring SNRs at least
as large as $10^2$ implies that resolving QNMs will be impossible for
redshifts larger than about 10.

The most probable outcome of the proposed no-hair test is a
verification of the BH hypothesis, combined with an accurate
measurement of the BH's mass and spin. This would be an exciting
experimental test of general relativity in the strong field
regime. However, we must leave the door open for surprises. 

What if {\em LISA} observations yield QNM frequencies that are not
compatible with the BH hypothesis? There are two possibilities:
either general relativity is not the correct theory of gravity, or the
compact object we are observing is not a black hole. If we
conservatively exclude the first hypothesis, the most plausible,
theoretically motivated alternative to explain the observations of
galactic centers is a boson star \cite{torres}. In the following we
present a bird's eye view on the theoretical properties of boson
stars, and propose a general-purpose test to ``count the hair'' of
supermassive objects. We also suggest that the hypothetical detection
of such a ``hairy'' object could provide interesting information on
particle physics and cosmology.

\subsection*{Exotic strawmen: boson star hair counting}  

Scalar particles such as the Higgs boson, the axion and the dilaton
play an important role in early Universe and high-energy
physics. Fundamental scalar fields have not been detected yet, but if
they exist they could condense to form boson stars: localized
solutions of the Einstein-Klein-Gordon system which are a natural
generalization of Wheeler's geons \cite{RB}.

Boson stars are macroscopic quantum states prevented from complete
gravitational collapse by Heisenberg's uncertainty principle.  In this
sense they are ``gravitational atoms'' that can have macroscopic size
and large masses. The main difference between boson star models
consists in the scalar self-interaction potential \cite{SM}. To
simplify\footnote{More general self-interaction potentials have been
considered in \cite{genBS}. There are also non-singular,
time-dependent equilibrium configurations of self-gravitating real
scalar fields known as {\em oscillatons} \cite{oscillatons}.}, we can
distinguish between three broad classes of boson star models:

1) {\it Miniboson stars.} If the scalar field is {\it non-interacting}
\cite{RB}, the maximum boson star mass $M_{\rm max}\simeq 0.633
m^2_{\rm Planck}/m$ is much smaller than the Chandrasekhar mass for
fermion stars $M_{\rm Ch}\sim m^3_{\rm Planck}/m^2$ (hence the name).
To support supermassive objects we need an ultralight boson of mass
$m=8.45\times 10^{-26} {\rm GeV}~(10^6 M_\odot/M_{\rm max})$.

2) {\it Boson stars.} The requirement of ultralight bosons to have
astrophysical-size objects can be lifted if we consider {\it
  self-interacting} scalar fields with a quartic self-interaction
potential of the form $\lambda |\phi|^4/4$ \cite{CSW}. As long as the
coupling constant $\lambda \gg (m/m_{\rm Planck})^2$ the maximum boson
star mass can be of the order of the Chandrasekhar mass or larger,
$M_{\rm max}\simeq 0.062 \lambda^{1/2} m^3_{\rm Planck}/m^2$.
Supermassive objects can exist if the boson mass and coupling constant
$\lambda$ are such that $m=3.2\times 10^{-4}{\rm GeV}~\lambda^{1/4}
(10^6 M_\odot/M_{\rm max})^{1/2}$.

3) {\it Nontopological soliton stars.} If the self-interaction takes
the form $U=m^2|\phi|^2(1-|\phi|^2/\phi_0^2)^2$ (where $\phi_0$ is a
constant) we can have nondispersive solutions with a finite mass,
confined to a finite region of space, even in the absence of gravity
\cite{LP}. The critical mass $M_{\rm max}\simeq 0.0198m^4_{\rm
  Planck}/(m \phi_0^2)$. One usually assumes $\phi_0\sim m$, so that a
$\sim 10^6 M_\odot$ boson star corresponds to a relatively heavy boson
with $m\sim 500$~GeV.

Boson stars are good ``strawmen'' for supermassive BHs. They are
indistinguishable from BHs in the Newtonian regime. Boson stars being
very compact, deviations in the properties of orbiting objects occur
close to the Schwarzschild radius and are very hard to detect
electromagnetically \cite{torres}. If the scalar field interacts only
gravitationally with matter compact objects could safely inspiral {\it
inside} the boson star, the only basic difference with a BH being the
absence of an event horizon \cite{kesden}. Boson masses yielding
supermassive boson star candidates differ by so many orders of
magnitude that even if we could set independent bounds on $m$ (by,
say, high-energy particle collisions) some candidates would still
survive.

The main point we wish to make is that {\it ringdown waves could
provide a way to unambiguously rule out boson stars as supermassive BH
``strawmen''.} If instead (serendipitously) gravitational wave
observations should turn out to be compatible with boson stars, we
could learn something about scalar field masses and their interaction
with matter, perhaps shedding some light on the low-energy limit of
string theories.

Any credible alternative to astrophysical Kerr BHs must include the
effects of rotation.  Stable rotating boson star solutions with
$\lambda \gg (m/m_{\rm Planck})^2$ were obtained by Ryan
\cite{ryan}. Considering a {\it pure state} for the scalar field of
the form $\phi=\Phi(r,\theta) e^{i(s\varphi-\Omega t)}$, Ryan showed
that the structure of a rotating boson star is determined by three
independent parameters: 1) a scaling factor $\lambda^{1/2}/m^2$, 2)
$\tilde \Omega=\Omega/m$ and 3) $\tilde
s=s/(\lambda^{1/2}/m)$\footnote{A sum over pure states would yield
time-dependent stress-energy tensors, so it's reasonable to expect
gravitational wave emission to lead the star to a pure state on short
timescales. The resulting configurations are torus-shaped, so one
should rather talk about ``boson doughnuts''. Notice also that the
uniqueness of the scalar field under a $2\pi$-rotation requires $s$ to
take only integer values, i.e. the angular momentum is {\it
quantized}. Despite this Ryan considers $\tilde s$ as a continuous
parameter, since the spacing of allowed values is extremely small.}.
Ryan's numerical structure calculations indicate that these three
parameters are in one-to-one correspondence with the first three
multipole moments of the star: mass $M$, spin $S_1$ and quadrupole
moment $M_2$\footnote{Interestingly, for a relatively slowly spinning
boson star ($S_1/M^2\lesssim 0.2$) the dimensionless multipole moments
are {\em almost constant}, and much larger than the value of unity
corresponding to a Kerr BH: $10\lesssim -M_2 M/S_1^2\lesssim 150$,
$20\lesssim -S_3 M/S_1^3\lesssim 200$. Low masses, corresponding to
less compact stars, have larger multipole moments. This weak
variability as a function of the rotation rate could be a useful
feature for our ``hair counting'' proposal.}. Given a measurement of
these moments one can infer the three boson star parameters. A fourth
measurement can then be used to test the boson star hypothesis, which
is really a {\it ``three-hair'' theorem} (in the same way as the BH
no-hair test is really a ``two-hair'' test).

The extension of the Kerr BH ``two-hair test'' to a ``hair counting
test'' for rotating supermassive objects requires the knowledge of QNM
frequencies for rotating boson stars with self-interaction of the type
considered by Ryan. Unfortunately this calculation has not been
performed yet, but some insight on the feasibility of the test can be
obtained from the QNM frequencies of nonrotating miniboson stars,
which are known \cite{YEF}. 

For the dominant quadrupole perturbations ($l=2$) the resulting QNM
spectrum is {\it very different from the QNM spectrum of Kerr
BHs}\footnote{In fact, the structure of the spectrum is reminiscent of
the $w$-modes of relativistic stars.}.  The ringdown signal from boson
stars should be a superposition of many modes with comparable
damping. For all modes the imaginary part is about an order of
magnitude smaller than the real part (for Kerr BHs this happens for
modes with $m>0$ in the limit $a/M\to 1$). The SNR for different modes
should be comparable, requiring a multi-mode data analysis similar to
the one developed in \cite{BCW}.

These expectations are confirmed by recent three-dimensional
time-evolutions of perturbed boson stars.  Fig.~4c of
Ref.~\cite{BBDGS} shows that many modes are indeed present in the
signal, that the $l=2$ component dominates the energy emission, and
that all modes with $n<12$ are necessary to fit the waveform (modes
with $n=5-10$ being dominant).

To assess detectability by {\em LISA}, let us consider the modes with
$n=7-8$ (the most excited in the numerical simulations) as
representative. These modes have a dimensionless frequency
$\sigma=\omega/m\simeq 2.3+0.13i$ \cite{YEF}, yielding
%
\be 
f\simeq
4.7\times 10^{-2} (10^6 M_\odot/M_{\rm max})~{\rm Hz}\,,
\qquad
\tau\simeq 60~(M_{\rm max}/10^6~M_\odot)~{\rm s}\,.
\ee
These numbers should be compared with the typical supermassive BH
ringdown frequency (\ref{BHs}). Our simple estimate shows that simply
choosing the boson mass $m$ to match the {\em mass} of a galactic BH
also brings the QNM {\em frequencies} in the interesting range for
{\it LISA}.

Gravitational wave emission from oscillating boson stars with
self-interaction has not been studied so far, but some general
features can be anticipated from existing calculations of the {\em
radial} pulsation frequencies. Fig.~4a in \cite{SS2} shows that radial
frequencies for $\lambda\neq 0$ have the same order of magnitude as in
the non-interacting case, generally decreasing with $\lambda$ and
scaling as $\lambda^{-1/2}$ for large $\lambda$. Eventually an
explicit calculation for rotating, self-interacting boson stars of the
kind considered by Ryan (as well as for other models, including
soliton stars and oscillatons) will be necessary to perform our
proposed test\footnote{We mention in passing that Ref.~\cite{FG}
provides Newtonian estimates for the gravitational wave emission from
the {\em decay of excited states} of a newly formed boson star into
the (stable) ground state. Their mechanism is quite different from the
nonradial oscillations of a boson star in the ground state, but it is
interesting to note that their estimate for the gravitational wave
frequencies lies in the {\em LISA} range.}.

If boson stars exist and ringdown waves from these object will be
detected, the payoff could be great. In Ryan's model, the measurement
of two QNMs (four observables) should be enough to determine all three
boson star parameters {\em and} to assess the boson star nature of the
object. In particular, a measurement of $\lambda^{1/2}/m^2$ could be
important for high-energy physics (eg. to test the low-energy limit of
string theories). If instead ringdown frequencies are consistent with
a black hole, we would have a striking confirmation of Einstein's
general relativity in the strong-field regime.

\bigskip
\bigskip
\centerline{\bf Acknowledgments}
\bigskip

We thank Clifford Will for useful discussions.  VC acknowledges
financial support from the Funda\c c\~ao Calouste Gulbenkian through
Programa Gulbenkian de Est\'{\i}mulo \`a Investiga\c c\~ao
Cient\'{\i}fica. This work was supported in part by the National
Science Foundation under grant PHY 03-53180.

\bibliography{paper}

\end{document}